\documentclass[letterpaper,titlepage,11pt]{article}
\usepackage{hyperref}
\usepackage{amssymb,amsmath,amsfonts,enumerate}
\usepackage{graphicx}
\usepackage{boxedminipage}
\usepackage{bm}

\def\clock{{\count0=\time
           \divide\count0 60
           \ifnum\count0<10 0\fi\the\count0
           \multiply\count0 -60 \advance\count0 \time
           :\ifnum\count0<10 0\fi \the\count0
         }}
\newcommand{\timestamp}{{\small\vbox{\hbox{\tt\jobname.tex}
\hbox{\the\day/\the\month/\the\year, \clock}}}}

\newcommand{\beq}{\begin{equation}}
\newcommand{\eeq}{\end{equation}}
\newcommand{\ben}{\begin{displaymath}}
\newcommand{\een}{\end{displaymath}}
\newcommand{\beqa}{\begin{eqnarray}}
\newcommand{\eeqa}{\end{eqnarray}}
\newcommand{\bea}{\begin{eqnarray}}
\newcommand{\eea}{\end{eqnarray}}
\newcommand{\bean}{\begin{eqnarray*}}
\newcommand{\eean}{\end{eqnarray*}}
\newcommand{\ba}{\begin{array}}
\newcommand{\ea}{\end{array}}
\newcommand{\bi}{\begin{itemize}}
\newcommand{\ei}{\end{itemize}}

\numberwithin{equation}{section}

\begin{document}

\begin{titlepage}
\begin{flushright}
\end{flushright}
\vskip 2.cm
\begin{center}
{\bf\LARGE{All Static Black Holes in AdS$_3$}}
\vskip 1.5cm

{\bf Nidal Haddad
}
\vskip 0.5cm
\medskip
\textit{Department of Physics, Birzeit University}\\
\textit{P.O.Box 14, Birzeit, West Bank, Palestine}\\

\vskip .2 in
\texttt{ haddad.nidal02@gmail.com}

\end{center}

\vskip 0.3in

\baselineskip 16pt
\date{}

\begin{center} {\bf Abstract} \end{center} 

\vskip 0.2cm 

\noindent

In this work we find the general static vacuum solution of three dimensional gravity with negative cosmological constant. Even though all solutions are locally diffeomorphic to pure $AdS_{3}$, solutions that differ globally from the latter space exist. New solutions with black holes on the $AdS_3$ boundary are found in both global and Poincare coordinates. In the Poincare coordinates such solutions are known as black funnels and black droplets. The black funnel provided by our general static metric is dual to the Hartle-Hawking state in the $1+1$ boundary theory.
\end{titlepage} \vfill\eject

\setcounter{equation}{0}

\pagestyle{empty}
\small
\normalsize
\pagestyle{plain}
\setcounter{page}{1}

\newpage

\section{Introduction}
 
The general theory of relativity simplifies significantly in three dimensions and so the search for exact solutions becomes less difficult compared to higher dimensions. The reason behind this simplification is that three dimensional gravity has no dynamical degrees of freedom - there are no gravitational waves - and hence matter sources can not influence the local geometry around them, but they can, nevertheless, have global effects on the geometry \cite{Deser:1984a,Brown:1988a}. In three dimensions with zero cosmological constant, the geometry outside matter sources is flat, but yet, a point-like matter source, for instance, will make the global geometry that of cone \cite{Deser:1984a}. A similar thing happens when the cosmological constant is non-vanishing; the space is locally dS$_3$ or AdS$_3$ for positive or negative cosmological constant respectively, but the spaces can be different globally (e.g., see \cite{Brown:1988a,Carlip:1995zj,Carlip:1995qv,Mann:1995eu}).   

In this paper we focus on three dimensional gravity with negative cosmological constant, and we find the general static vacuum metric. The general solution includes metrics that differ globally from AdS$_3$ by containing \textbf{horizons}. 
The general solution includes a very interesting family of black hole solutions. Those black holes can be viewed as extended black objects embedded in global AdS$_3$. More precisely, they describe horizons that extend from the boundary of AdS$_3$ down into the bulk to encircle a Schwarzschild-AdS black hole located in the center (see Fig.[\ref{fig:FIG1}]). In Poincare coordinates the latter solution is viewed as a black hole extending from the boundary of AdS$_3$ down into the bulk, and connecting (or merging) with a planar black hole that is located in the infrared (see the left figure in Fig.[\ref{fig:FIG0}]); it is the so-called black funnel solution \cite{Hubeny:2009ru,Hubeny:2009kz}.
The general solution includes also another interesting class of solutions; a black hole that dangles from the boundary of pure AdS$_3$ and closes off at some point in the bulk (see Fig.[\ref{fig:FIG2}]). This solution could be termed a black droplet, although, in general, a black droplet solution contains a central bulk black hole in addition to the droplet.

One of the interesting features of those solutions is that they induce a black hole metric on the AdS$_{3}$ boundary. According to the AdS/CFT correspondence, \cite{Maldacena:1998a,Aharony:1999ti}, those $3-$dimensional gravitational solutions are dual to $1+1$ conformal field theories living on curved backgrounds. In other words, they are dual to hawking radiation from $2-$dimensional black holes at large N and at strong coupling (see \cite{Hubeny:2009ru,Hubeny:2009kz,Fischetti:2012ps} for details). As we will see in the bulk of the paper the black funnel solution provided by our general static metric is dual to the Hartle-Hawking state; the boundary black hole and the plasma are in thermal equilibrium.   

The paper is organized as follows. We start in section \ref{sec:fun} by giving a very short introduction to strongly coupled field theories in black hole backgrounds and their gravitational duals - black funnels and black droplets. In section \ref{sec:basic} we introduce the basic set up and the main results. We give the general static metric in AdS$_3$, we analyze the possible horizon shapes, and then we focus on the black funnel solution. In section \ref{sec:boundary} we focus on the $1+1$ boundary metric of the black funnel, and  we compute its holographic stress tensor. We give the details of the derivation of the general solution in appendix \ref{app:derivation} . 

\section{Black funnels and black droplets}
\label{sec:fun}

\begin{figure}[ht]
\centering
\begin{minipage}[b]{0.45\linewidth}

\includegraphics[bb=0 0 250 180,scale=0.78]{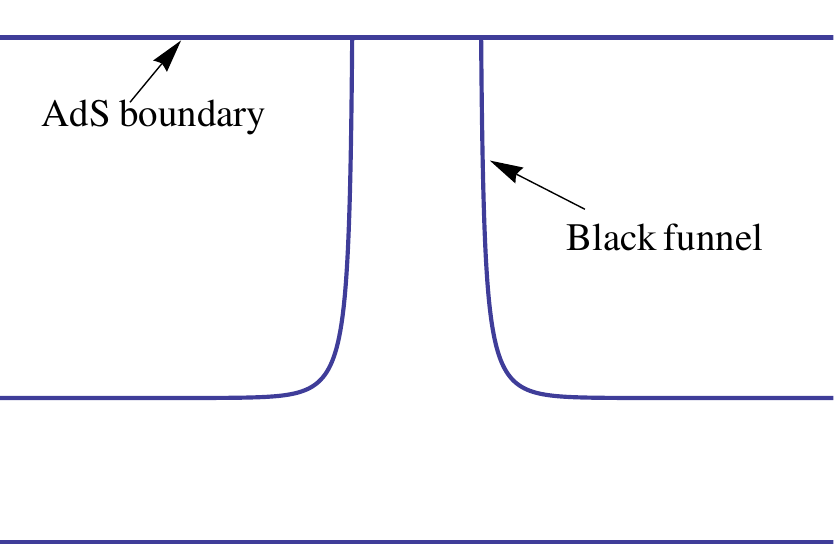}
\label{fig:minipage01}
\end{minipage}
\quad
\begin{minipage}[b]{0.45\linewidth}

\includegraphics[bb=0 0 250 180,scale=0.78]{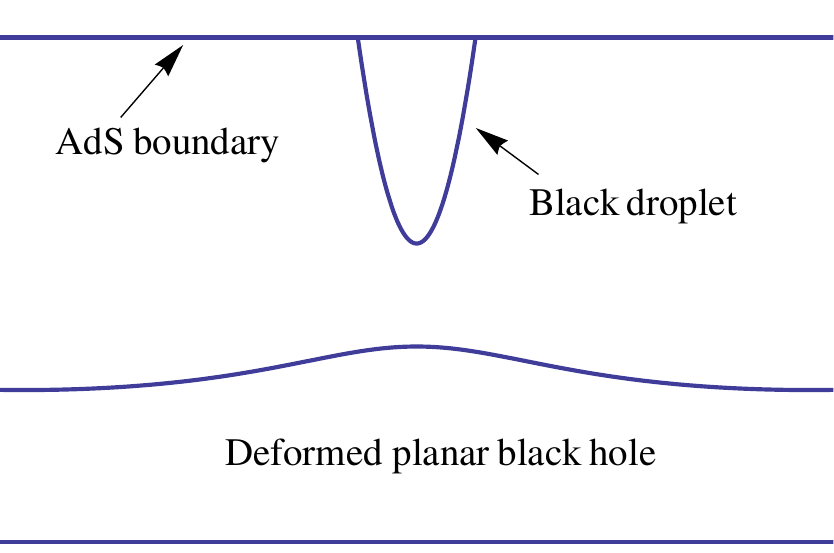}
\label{fig:minipage02}
\end{minipage}

\caption{\small Left: This is a black funnel. Note how the horizon extends from the boundary of AdS and merge with the planar black hole in the bottom in a shoulder-like configuration. Right: This is a black droplet. The horizon extends from the boundary of AdS downwards and it caps off before it reaches the planar black hole which gets a slight deformation.
 }

\label{fig:FIG0}
\end{figure}

In the last few decades, the study of quantum fields in black hole backgrounds focused on weakly coupled fields (see \cite{Birrell:1982a,Jacobson:2003vx} for reviews). Very little was known about strongly coupled fields. Fortunately, after the advent of the AdS/CFT correspondence \cite{Maldacena:1998a,Aharony:1999ti} it has become possible to study and learn about strongly coupled quantum field theories as the AdS/CFT correspondence constructs a one-to-one map between weakly coupled theories of gravity on AdS spacetimes and strongly coupled conformal field theories living on the boundary of AdS. For pure AdS for example the induced metric on the AdS boundary is Minkowski and thus one would say that the field theory is living in flat space. In fact, one can use the AdS/CFT correspondence to study the more general situation of strongly coupled fields on black hole backgrounds and that is by looking for black hole solutions in AdS with horizons that extend to the boundary, thus inducing a black hole metric on the boundary of AdS where the dual field theory lives. Two types of such solutions were conjectured to exist \cite{Hubeny:2009ru}; black funnels and black droplets. Black funnels are black holes with connected horizons that extend from the boundary to horizon of the planar Schwarzschild-AdS geometry - they connect with the planar black hole horizon in a shoulder-like configuration (see Fig.[\ref{fig:FIG0}]). Black droplets, on the other hand, are black holes with disconnected horizons, that is, they extend from the boundary of AdS down to some point in the bulk where they close off (or cap off) in a smooth way before they reach the planar black hole horizon; the planar black hole gains some deformation as a result of the droplet suspended above it (see Fig.[\ref{fig:FIG0}]). Black funnels and droplets are the gravitational duals of different vacuum states of $\mathcal{N}=4$, $SU(N)$ super Yang-Mills theory on black hole backgrounds, at large $N$ and large 't Hooft coupling.
There are some physical differences between them though. Black funnels (as the horizon is connected) are dual to a deconfined plasma which is strongly coupled to the boundary black hole, that is, energy trasfer is quick between them, of order $O(N^2)$. Black droplets on the other hand (as the two horizons in the bulk are disconnected) are dual to a deconfined plasma which is coupled weakly to the boundary black hole, that is, energy transfer is slow between them, of order $O(1)$ (see \cite{Hubeny:2009ru,Fischetti:2012ps} for further details).
In general, the temperature of the boundary deconfined plasma can be different from the temperature of the boundary black hole, depending on the sizes of the planar black hole and the boundary black hole, respectively. If the two temperatures are equal this is dual to the Hartle-Hawking vacuum state, describing thermal equilibrium betweem the plasma and the boundary black hole (in this paper and in \cite{Hubeny:2009ru} there are examples of such states). If the two temperatures are different this is the dual of the Unruh or the Boulware vacuum states; those are out of equilibrium states (see \cite{Figueras:2011va,Haddad:2012ss,Haddad:2013tha} for examples).  

\section{Main Results}
\label{sec:basic}
The equations we are interested to solve in this paper are the Einstein equations with negative cosmological constant in three dimensions. Namely, 

\beq \label{EOM}
E_{\mu\nu}-\frac{1}{L^2}g_{\mu\nu}=0\,,
\eeq
where $E_{\mu\nu}\equiv R_{\mu\nu}-\frac{1}{2}Rg_{\mu\nu}$ is the Einstein tensor, and $L$ is the radius of curvature. One can rewrite the above equations as,

\beq
R_{\mu\nu}=-\frac{2}{L^2}g_{\mu\nu}\,.
\eeq
The known static (and non-singular) solutions of the above equations are pure AdS$_3$ and Schwarzschild-AdS$_{3}$. In global coordinates the Schwarzschild-AdS$_{3}$ metric is given by,
 
\beq
ds^2=-\left(r^2/L^2-M\right)dt^2+\frac{dr^2}{r^2/L^2-M}+r^2d\theta^2\,,
\eeq
where $M$ is the mass parameter of the black hole, and it is dimensionless in $3$ dimensions. This is the non-rotating BTZ black hole \cite{Banados:1992wn}. Note that the pure AdS$_{3}$ metric can be obtained from the Schwarzschild-AdS$_{3}$ metric by making the mass negative with the specific value $M=-1$ (see \cite{Banados:1992wn}),

\beq
ds^2=-\left(r^2/L^2+1\right)dt^2+\frac{dr^2}{r^2/L^2+1}+r^2d\theta^2\,.
\eeq

It is to be emphasized that the above solutions are spherically symmetric (the solutions depend only on the radial coordinate $r$) whereas the general solution we are going to give below is not spherically symmetric - it is the general static metric that depends on the two spatial coordinates $r$ and $\theta$.

\subsection{The General Static Solution}
\label{sec:General Solution}
The general static solution we have found is given by the metric (the derivation is given in appendix \ref{app:derivation}),
\beq\label{general}
ds^2=-\left(r^2/L^2-M\right)\left(A_0+\frac{A_1 e^{-\sqrt{M}\theta}+A_2 e^{\sqrt{M}\theta}}{\sqrt{1-ML^2/r^2}}\right)^2dt^2+\frac{dr^2}{r^2/L^2-M}+r^2d\theta^2\,,
\eeq
where $A_0$, $A_1$, and $A_2$ are arbitrary constants. One can clearly see that this metric contains a horizon as the component $g_{tt}$ can be made zero along some specific contour $r(\theta)$. Note that if the constant $A_0$ is not zero then it can be set to one, $A_0=1$, by a rescaling of the time coordinate. The physical meaning of the constants $A_1$ and $A_2$ is that they (as we will see in details as we proceed) play a role in determining the location of the horizon. Note, furthermore, that this solution must be required to be periodic in the coordinate $\theta$. It worth noting as well that the Schwarzschild-AdS$_{3}$ black hole is obtained from the above general metric by taking $A_0=1$ and $A_1=A_2=0$.

\begin{figure}[ht]
\begin{center}
\includegraphics[bb=0 0 250 250,scale=0.9]{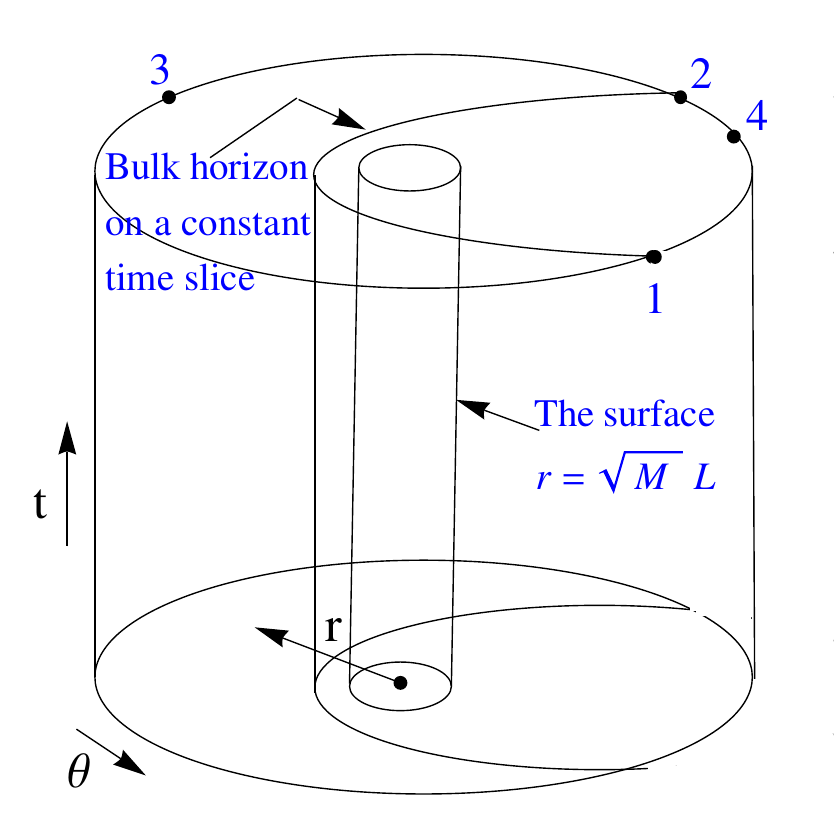}
\end{center}
\caption{\small The coordinates $\left(t,r,\theta\right)$ are shown in the figure. On a constant time slice the bulk horizon is pointed out. The dotted points 1 and 2 are the locations of the boundary black hole horizon. The point 3 lies on the AdS$_3$ boundary outside the black hole, while the point 4 lies inside. The surface $r=\sqrt{M}L$ (the cylinder) is also specified in the figure.
It is worth to stress the generic feature that the bulk horizon never touches the surface $r=\sqrt{M}L$.
}\label{fig:FIG1}
\end{figure}

Now we will focus on the case with non-vanishing $A_0$ (so we set $A_0=1$) and we will analyze the following solution,
\beq\label{general'}
ds^2=-\left(r^2/L^2-M\right)\left(1-\frac{e^{\mp\sqrt{M}(\theta\mp\theta_0)}}{\sqrt{1-ML^2/r^2}}\right)^2dt^2+\frac{dr^2}{r^2/L^2-M}+r^2d\theta^2\,,
\eeq
where the upper sign refers to the range $0\leq\theta\leq\pi$ while the lower sign refers to the range $-\pi\leq\theta\leq 0$. Notice that we have chosen the constants $A_1$ and $A_2$ so as to make the solution periodic in $\theta$ (as it should) and symmetric around $\theta=0$. For $\theta\in[0,\pi]$ we have taken $A_1=-e^{\sqrt{M}\theta_0}$, $A_2=0$ and for $\theta\in[-\pi,0]$ we have taken $A_1=0$, $A_2=-e^{\sqrt{M}\theta_0}$. It is important to notice that $\theta=\pm\theta_0$ are the locations of the boundary horizon since for $r\rightarrow\infty$ the component $g_{tt}$ vanishes there. The location of the (bulk) horizon is obtained  by solving the equation $g_{tt}=0$. For the metric (\ref{general'}) the location of the (bulk) horizon is given by the equation,
\beq
r_{H}(\theta)=\frac{\sqrt{M}L}{\sqrt{1-e^{\mp 2\sqrt{M}(\theta\mp\theta_0)}}}\,,
\eeq  
and it is illustrated in Fig.[\ref{fig:FIG1}] which shows how this space is embedded in global $AdS_3$. The alert reader will note that the metric (\ref{general'}) exhibits a discontinuity in the derivative at $\theta=0$; this singularity, however, is not a problem as it is hidden inside the horizon.   

\begin{figure}[ht]
\begin{center}
\includegraphics[bb=0 0 250 250,scale=0.9]{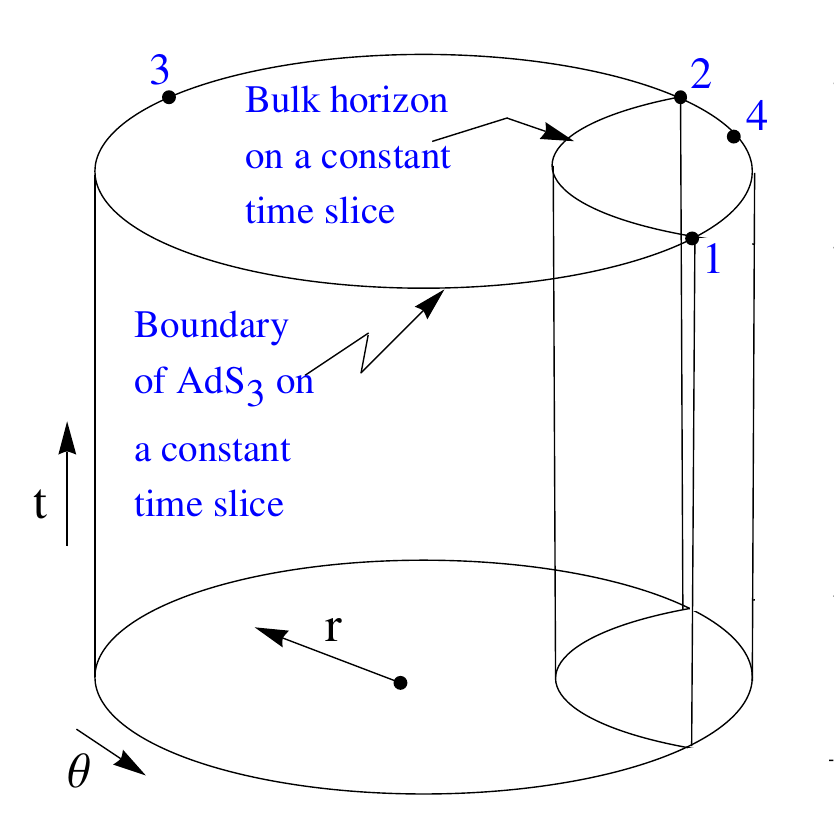}
\end{center}
\caption{\small On a constant time slice the bulk horizon and the boundary of $AdS_3$ are pointed out. The points 1 and 2 are the locations of the boundary black hole horizon. The point 3 lies outside the horizon, while the point 4 lies inside.  
}\label{fig:FIG2}
\end{figure}

Another interesting subcase to stop at is the case with $M=-1$, which reads,
\beq\label{droplet}
ds^2=-\left(r^2/L^2+1\right)\left(A_0+\frac{a_1\cos\theta+a_2\sin\theta}{\sqrt{1+L^2/r^2}}\right)^2dt^2+\frac{dr^2}{r^2/L^2+1}+r^2d\theta^2\,,
\eeq
where $a_1$ and $a_2$ are arbitrary constants (they are linear combinations of the arbitrary constants $A_1$ and $A_2$). Here also the metric contains a horizon as the component $g_{tt}$ can be made zero along some specific contour $r(\theta)$. Next we will focus on the case with $A_0=1$. Without loss of generality we can take $a_2=0$, \footnote{The expression $a_1\cos\theta+a_2\sin\theta$ is equal to $a\cos(\theta+b)$ for some $a$ and $b$. The constant $b$ corresponds to a shift in the coordinate $\theta$ (it corresponds to the location of $\theta=0$ along the circle) and so we can set it to zero.} 
\beq\label{droplet'}
ds^2=-\left(r^2/L^2+1\right)\left(1+\frac{a_1\cos\theta}{\sqrt{1+L^2/r^2}}\right)^2dt^2+\frac{dr^2}{r^2/L^2+1}+r^2d\theta^2\,.
\eeq
The location of the (bulk) horizon is obtained  by solving $g_{tt}=0$. For the metric (\ref{droplet'}) the horizon's location is given by the equation:
\beq
r_H=\frac{L}{\sqrt{a_1^2\cos^2\theta-1}}\,,
\eeq
provided that $a_1\geq1$ (otherwise there is no horizon). See Fig.[\ref{fig:FIG2}] which shows how this space is embedded in global $AdS_3$ and to see how the horizon looks like. This solution can be viewed as pure AdS$_3$ with a black hole that extends from the boundary and closes off at some point in the bulk - a black droplet in pure AdS$_3$.

It is worth to mention briefly at this point the case $A_0=0$. One can easily check that then the solutions (\ref{general}) and (\ref{droplet})
are characterized by having boundary metrics of constant negative/positive curvature respectively.

Finally, we would like to mention that the black hole solutions provided above have no singularities inside their horizons; recall that the same is true for the BTZ black hole for it has no singularity at $r=0$ (e.g., see \cite{Carlip:1995qv}).

\subsection{Black Funnels}
\label{subsec:Black Funnels}
\begin{figure}[ht]
\centering
\begin{minipage}[b]{0.45\linewidth}

\includegraphics[bb=0 0 250 180,scale=0.78]{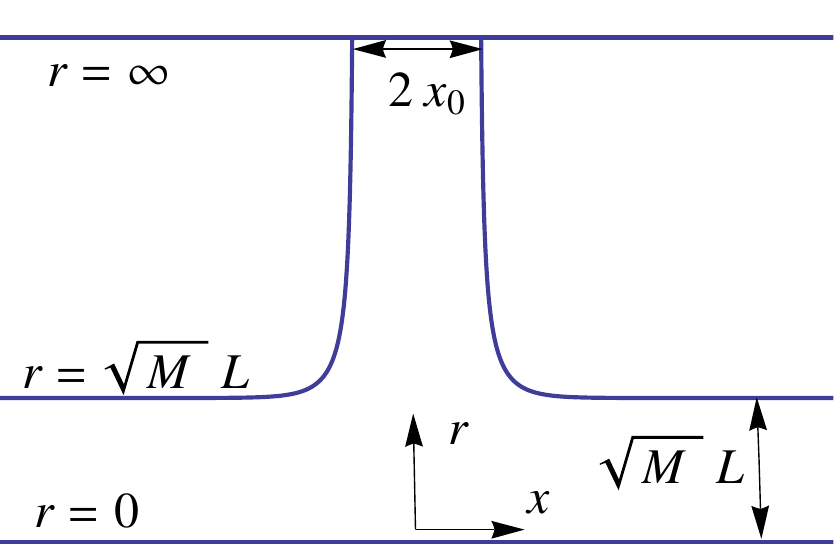}
\label{fig:minipage1}
\end{minipage}
\quad
\begin{minipage}[b]{0.45\linewidth}

\includegraphics[bb=0 0 250 180,scale=0.78]{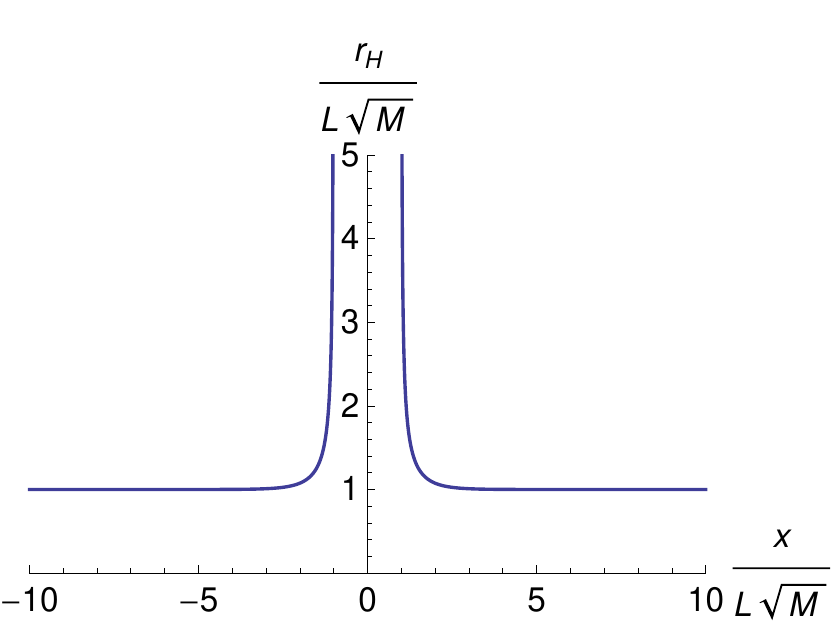}
\label{fig:minipage2}
\end{minipage}

\caption{\small Left: This is a black funnel. $r=\infty$ is the AdS$_3$ boundary. $x_0$ and $M$ are free parameters. $2x_0$ is the size of the boundary black hole. $M$ is the mass parameter of the planar black hole which the funnel reduces to at large $\left|x\right|$. Right: This is a normalized plot of the function $r_H(x)$ - see eq.(\ref{funnel}) - describing the location of the horizon in the $r-x$ plane. We have plotted $\frac{r_H}{L\sqrt{M}}$ as a function of the normalized coordinate $\frac{x}{L\sqrt{M}}$. Note that $r_H\rightarrow\infty$ at $x=\pm x_0$ where in this figure we have chosen $x_0=\sqrt{M}L/2$ for the sake of illustration only.
 }

\label{fig:FIG3}
\end{figure}

To connect this story to black funnels \cite{Hubeny:2009ru,Hubeny:2009kz} take the metric (\ref{general'}) and uncompactify the $\theta$ coordinate. 	That is, define a new coordinate $x$ by $x=L\theta$ and let the range of $x$ be $(-\infty,\infty)$. The metric that we will get is, 
\beq\label{black funnel}
ds^2=\frac{L^2dr^2}{r^2\left(1-ML^2/r^2\right)}+\frac{r^2}{L^2}\left[-\left(1-ML^2/r^2\right)\left(1-\frac{e^{\mp\sqrt{M}(x\mp x_0)/L}}{\sqrt{1-ML^2/r^2}}\right)^2dt^2+dx^2\right]\,,
\eeq
where $x_0\equiv L\theta_0$ and here again the upper and lower signs refer to $x\geq0$ and $x\leq0$, respectively. We are going to argue next that this metric describes a black funnel. Note first that this metric reduces to the planar black hole metric for large $\left|x\right|$ as is expected from a black funnel. What makes it obvious that it is a funnel is the shape of the horizon,
\beq\label{funnel}
r_H=\frac{\sqrt{M}L}{\sqrt{1-e^{\mp 2\sqrt{M}(x\mp x_0)/L}}}\,,
\eeq 
which is manifestly of a funnel shape (see plots in Fig.[\ref{fig:FIG3}]). The alert reader will note that the metric (\ref{black funnel}) exhibits a discontinuity in the derivative at $x=0$; this singularity, however, is not a problem as it is hidden inside the horizon.   

To understand better the thermodynamic aspects of the black funnel we find it useful to compute its temperature. We find its temperature (computed on constant $r$ slices) to be,
\beq
T=\frac{\sqrt{M}}{2\pi L}\sqrt{1-ML^2/r^2}\,.
\eeq
The temperature of the boundary black hole is obtained by taking the limit $r\rightarrow\infty$, upon which it becomes,
\beq\label{temp}
T=\frac{\sqrt{M}}{2\pi L}\,,
\eeq
which, interestingly, equals exactly the temperature of the planar black hole. Physically this means that there is a thermodynamic equilibrium between the boundary black hole and the planar black hole in the bulk; in special there is no heat flow in the funnel from the boundary to the bulk or vice versa. From the boundary theory point of view this is viewed as a thermal equilibrium between the plasma and the boundary black hole, or, in other words, this is seen as the Hartle-Hawking state \cite{Hubeny:2009ru,Fischetti:2012ps}. This last point explains the reason behind the fact that the temperature (\ref{temp}) does not depend on the size of the boundary black hole $x_0$; to have a static solution there must be no heat flow between the plasma and the boundary black hole, and so their temperatures must be equal. If the temperature of the boundary black hole were different from that of the plasma (that is, if it depended on $x_0$) then there would be a heat flow between the two, and in special this would break staticity (this would be an out of equilibrium state such the Unruh state).

\section{Boundary Metric and Holographic Stress Tensor}
\label{sec:boundary}
In this section we are going to focus on the black funnel metric (\ref{black funnel})\footnote{The same treatment applies equally to the metric (\ref{general'}) since the two metrics are related by the simple replacement $x=L\theta$}. One can easily check that the boundary metric for the black funnel is,
\beq
ds_{1+1}^2=-\left(1-e^{\mp\sqrt{M}(x\mp x_0)/L}\right)^{2}dt^2+dx^2\,.
\eeq
Obviously this is an asymptotically flat metric with a horizon at $x=\pm x_0$. The boundary stress tensor is readily computed from the metric (\ref{black funnel}) using the prescription of \cite{Balasubramanian:1999re}\footnote{Our notations introduce a minus sign difference from \cite{Balasubramanian:1999re} in $\Theta_{ab}$. We use the notations of \cite{Wald:1984}, not of \cite{Weinberg:1972}, and therefore we have a minus sign difference in the Riemann tensor with respect to \cite{Balasubramanian:1999re,Weinberg:1972}.}, 
\begin{equation}\label{}
T_{ab}=\frac{1}{8\pi G_{3}}\left[-\Theta_{ab}+\Theta\gamma_{ab}-\frac{1}{L}\gamma_{ab}\right]\,,
\end{equation}   
where $\gamma_{ab}$ is the induced metric on an $r=\text{constant}$ surface, $\Theta_{ab}=(\nabla_{a}n_{b}+\nabla_{b}n_{a})/2$ is the extrinsic curvature of that surface, where $n_{a}$ is an outward pointing normal vector to the surface. The non-vanishing components of the boundary stress tensor turn out to be,
\beq
T_{tt}=\frac{M}{16\pi G_{3}L}\left(1-e^{\mp\sqrt{M}(x\mp x_0)/L}\right)^{2}\,
\text{,}\,\qquad
T_{xx}=\frac{M}{16\pi G_{3}L}\coth\left(\frac{\sqrt{M}(x\mp x_0)}{2L}\right)\,.
\eeq
As a first check, note that for $x\rightarrow\pm\infty$ the stress tensor reduces to the one of a planar black hole as it should, that is, it reduces to $T_{tt}=T_{xx}=\frac{M}{16\pi G_{3}L}$. This is an expected result since far away from the neck of the funnel ($x\rightarrow\pm\infty$) the spacetime reduces to that of a planar black hole. On the other hand, for $x\rightarrow \pm x_0$, we see that,
\beq
T_{tt}\rightarrow 0\,
\text{,}\,\qquad
T_{xx}\rightarrow\infty\,,
\eeq
and so we see that there is a diverging component in the stress tensor. The trace of the stress tensor is,
\beq
T^\mu_\mu=\gamma^{\mu\nu}T_{\mu\nu}=\frac{M}{8\pi G_{3}L}\left(\frac{1}{e^{\pm\sqrt{M}(x\mp x_0)/L}-1}\right)\,,
\eeq
and as can be easily seen the trace diverges as $x\rightarrow \pm x_0$; therefore, we conclude that the divergency of the stress tensor at the boundary black hole horizon is coordinate independent. The interpretation of this divergency from the CFT point of view is left for future work. Regarding the trace anomaly, one can check that the Ricci scalar of the boundary metric is,
\beq\label{2DimRicci}
R=\frac{2M}{\left(e^{\pm\sqrt{M}(x\mp x_0)/L}-1\right)L^2}\,,
\eeq
and hence the stress tensor satisfies $T^\mu_\mu=\frac{c}{24\pi}R$ with $c=\frac{3L}{2G_3}$ as it should \cite{Balasubramanian:1999re}. \footnote{According to the conventions of \cite{Balasubramanian:1999re} $T^\mu_\mu=-\frac{c}{24\pi}R$ which introduces a minus sign difference from our result.}

Now we turn to discuss some points concerning the structure of the boundary metric. As can be seen from the Ricci scalar (\ref{2DimRicci}) there is a singularity at $x=\pm x_0$, that is, the black hole horizon of the boundary theory is singular. This however does not mean that there is something wrong since we know  that our bulk $3-$dimensional metric is regular with no singularities anywhere. In contrast, one notices from the Ricci scalar that there is no singularity inside the horizon as the curvature is finite there (also the metric is finite inside the horizon). This is similar to the BTZ black hole which has no singularity at $r=0$. Finally, in order to see clearly that the surface $x=\pm x_0$ is a horizon look at the limit $x\rightarrow \pm x_0$. This gives the Rindler space, $ds_{1+1}^2\approx-\frac{M}{L^2}(x\mp x_0)^2dt^2+dx^2$, which, as is well known, has a horizon at $x=\pm x_0$.

\section{Discussion}
\label{sec:discuss}

In this work we have obtained the general static metric in AdS$_3$. It is important to compare our results with those obtained in \cite{Skenderis:1999nb}. In the paper \cite{Skenderis:1999nb} the authors had found the general static metric in AdS$_3$ given an arbitrary boundary metric. In our work, however, the boundary metric is not arbitrary but it is restricted to a certain class of metrics. We conclude from this that the source of this difference is that the authors in \cite{Skenderis:1999nb} did not fix the gauge of their solution completely; the metric that they provided has a gauge redundancy in it, and if this remaining gauge freedom is fixed then the boundary metrics will be restricted to some class of metrics as in our work. Moreover, in the work \cite{Skenderis:1999nb} the authors did not discuss the horizon shapes of their black hole metrics as we did - we have illustrated how the horizons are embedded in global AdS$_3$ and we have discussed the important metrics describing black funnels. It is also to be said that our method of finding the general metric is different from  \cite{Skenderis:1999nb}. We have solved the Einstein equations directly while they have resorted to indirect methods to do so; they defined an auxiliary scalar field to recover covariance and they went to solve some type of a Liouville action. Finally, in the coordinate system we have worked with the metric, fortunately, turned out to have a compact and simple form, which makes it easy to use and analyze.

We also want to comment on relevant results and analysis made in \cite{Hubeny:2009ru,Fischetti:2012ps}. It is important to stress the following point: The fact that any solution to $3-$dimensional gravity with negative cosmological constant is locally $AdS_3$ does not mean that any solution to the latter can be obtained from $AdS_3$ by a coordinate transformation. There could be solutions that are globally different from $AdS_3$ by having horizons or by having conical singularities, and those can not be obtained from AdS$_3$ by a coordinate transformation. Thus the $3-$dimensional black funnels obtained in \cite{Hubeny:2009ru,Fischetti:2012ps} are not the most general funnels since they are obtained from the BTZ black hole by a coordinate transformation.

With some overlap here and there we think that our work together with the work of \cite{Skenderis:1999nb} and \cite{Hubeny:2009ru,Fischetti:2012ps} give a complete picture of the subject dealing with static metrics in $AdS_3$.

It is important also to say some words regarding the stability of the static solutions provided in this article. One may wonder if these solutions (and their horizons) are classically stable (for example, if the black funnel suffers from a Gregory-Laflamme instability \cite{Gregory:1993vy,Gregory:2000gf}). However, one recalls immediately that in three dimensional gravity there are no dynamical degrees of freedom and so any given static solution of the theory is guaranteed to be stable. In other words, if we perform the classical stability analysis by solving the linearized equations we will discover that there are no time-dependent modes and so all solutions are stable. Therefore, the static solutions provided in this article are classically stable.

We end the discussion by stressing that this paper is a simplified model, or an exercise in $(2+1)$-dimensional gravity intended to provide insights into its generalization to higher dimensions. Looking for funnels and droplets in higher dimensions is a non-easy task and that is why it is recommended as a first try to work in lower dimensions where gravity and also the CFT simplify significantly.

\subsection*{Acknowledgement}
It is a pleasure to thank Nabil Iqbal for very important discussions, for reading and for commenting on the draft.

\appendix

\section{Derivation}
\label{app:derivation}

We are interested to find the general static solution of the equations (\ref{EOM}). The general static metric in three dimensions can be put in the form (see Appendix \ref{app:gen} for proof),
\beq
ds^2=g_{rr}(r,x)dr^2+g_{tt}(r,x)dt^2+g_{xx}(r,x)dx^2\,,
\eeq
where $g_{rr}(r,x)$ is an arbitrary function. Without loss of generality and for reasons to be explained momentarily we choose to write our metric as,
\beq \label{general metric}
ds^2=\frac{L^2dr^2}{r^2\left(1-ML^2/r^2\right)}+\frac{r^2}{L^2}\left[-\left(1-ML^2/r^2\right)f(r,x)dt^2+g(r,x)dx^2\right]\,.
\eeq
The motivation for choosing the above general form is that it reduces to the well-known planar black hole metric when $f(r,x)=g(r,x)=1$ - and hence in special the holographic direction $r$ is pointed out. Furthermore, since we are interested in finding black funnel solutions this form is appropriate because black funnels are expected to reduce to the planar black hole metric in the regions $x\rightarrow\pm\infty$. 
\footnote{Note that the metric (\ref{general metric}) can be put in the Fefferman-Graham form $ds^2=(dz^2+g_{\mu\nu}(x,z)dx^\mu dx^\nu)/z^2$ by a reparameterization of $r$.}
\subsection{Field Equations}
There are 4 equations to be solved; the components $tt$, $xx$, $xr$, and $rr$ of the field equations (\ref{EOM}). The rest of the equations are automatically satisfied by our general form (\ref{general metric}).\\
\underline{\textbf{The $tt$ component:}}\\
The $tt$ component of the equations of motion (\ref{EOM}) reads,
\beq
\frac{\partial_r g}{g}-2\frac{\partial_r^2 g}{\partial_r g}=\frac{6r-4ML^2/r}{r^2-ML^2}\,.
\eeq
This equation can be easily integrated to give,
\beq\label{g}
g(r,x)=\left(F(x)\sqrt{1-ML^2/r^2}+G(x)\right)^2\,,
\eeq
where $F(x)$ and $G(x)$ are arbitrary functions of $x$.\\
\underline{\textbf{The $xx$ component:}}\\
The $xx$ component of the equations of motion (\ref{EOM}) reads,
\beq
\frac{\partial_r f}{f}-2\frac{\partial_r^2 f}{\partial_r f}=\frac{6r}{r^2-ML^2}\,,
\eeq
which can be easily integrated, in a similar way to the $tt$ equation, and its solution is

\beq\label{f}
f(r,x)=\left(\frac{A(x)}{\sqrt{1-ML^2/r^2}}+B(x)\right)^2\,,
\eeq
where $A(x)$ and $B(x)$ are arbitrary functions of $x$. \\
\underline{\textbf{The $xr$ component:}}\\
After inserting the above solutions for $f(r,x)$ and $g(r,x)$, the $xr$ component of the equations of motion gives,
\beq
F(x)A'(x)=G(x)B'(x)\,,
\eeq
which, if working carefully, must be divided into two cases.\\ 
\underline{\textbf{Case 1:}} If $B'(x)= 0$ then we can not divide by $B'(x)$ and so the $xr$ component reduces to,
\beq
F(x)A'(x)=0 \,
\text{}\,\qquad
\text{Case [1]}\,,
\eeq
which in turn also splits into two subcases, 
\beq
A'(x)=0 \,
\text{}\,\qquad
\text{Case [1a]}\,,
\eeq
and
\beq
F(x)=0 \,
\text{}\,\qquad
\text{Case [1b]}\,.
\eeq
\underline{\textbf{Case 2:}} If $B'(x)\neq 0$ then we can eliminate $G(x)$ by dividing the equation by $B'(x)$,
\beq
G(x)=\frac{F(x)A'(x)}{B'(x)}\,
\text{}\,\qquad
\text{Case [2]}\,.
\eeq
\underline{\textbf{The $rr$ component:}}\\
Insert the above solutions for $f(r,x)$ and $g(r,x)$ into the $rr$ component and get the following for the different cases.\\
For the case [1a]: Here we have $A'(x)=B'(x)= 0$, or equivalently, $A=A_0$ and $B=B_0$ where $A_0$ and $B_0$ are arbitrary constants. The $rr$ component of the equations of motion will read,
\beq
A_0G(x)=B_0F(x)\,.
\eeq
For the case [1b]: Here we have $B'(x)=F(x)=0$. Take $B=B_0$ where $B_0$ is an arbitrary constant. The $rr$ component of the equations of motion will read,
\beq \label{zz comp}
MAG^3+L^2\left(A'G'-A''G\right)=0\,.
\eeq
For the case [2]: Here we have $B'(x)\neq 0$ and $G(x)=F(x)A'(x)/B'(x)$ and so the $rr$ component of the equations will read,
\beq\label{eom1b}
MF^3AA'+B'\left[-MF^3B+L^2\left(B'F'-FB''\right)\right]=0\,. 
\eeq
The general solution comes from the case [1b] and from the case [2]. However, it is easier to obtain it from the case [1b], and that is what we will do next.
In appendix B we will derive the general solution from the case [2]. 
\subsection{The General Solution (Derived from the Case [1b])}
The general solution is derived from the Case [1b]. In this case, as mentioned previously, $B(x)=\text{constant}=B_0$ and $F(x)=0$, and the metric reads [see equations (\ref{general metric}),(\ref{g}),(\ref{f})],
\beq\label{metric1b}
ds^2=\frac{L^2dr^2}{r^2\left(1-ML^2/r^2\right)}+\frac{r^2}{L^2}\left[-\left(1-ML^2/r^2\right)\left(\frac{A(x)}{\sqrt{1-ML^2/r^2}}+B_0\right)^2dt^2+G(x)^2dx^2\right]\,. 
\eeq
There is one more equation to be satisfied still, the $rr$ component (\ref{zz comp}), which we repeat here for the sake of clarity, 
\beq \label{zz comp'}
MAG^3+L^2\left(A'G'-A''G\right)=0\,.
\eeq
Note from (\ref{metric1b}) that setting $G(x)=1$ is equivalent to a reparameterization of the $x$ coordinate and so we use this freedom and set 
\beq 
G(x)=1\,.
\eeq
Thereafter, (\ref{zz comp'}) reduces to the elementary O.D.E,

\beq 
A''=\frac{M}{L^2}A\,,
\eeq
whose general solution is,

\beq 
A(x)=A_1e^{-\sqrt{M}x/L}+A_2e^{\sqrt{M}x/L}\,,
\eeq
where $A_1$ and $A_2$ are the two integration constants. Thus, the general solution is,
\beq
ds^2=\frac{L^2dr^2}{r^2\left(1-ML^2/r^2\right)}+\frac{r^2}{L^2}\left[-\left(1-ML^2/r^2\right)\left(B_0+\frac{A_1e^{-\sqrt{M}x/L}+A_2e^{\sqrt{M}x/L}}{\sqrt{1-ML^2/r^2}}\right)^2dt^2+dx^2\right]\,,
\eeq
where $B_0$, $A_1$ and $A_2$ are arbitrary constants. The constants $A_1$ and $A_2$ are to be fixed by boundary conditions. As for the constant $B_0$ there are two possibilities. The first possibility is $B_0=0$\footnote{One can check that in this case the solution is characterized by a boundary metric with constant negative curvature.}, and the second is $B_0\neq0$ in which case by a rescaling of the time coordinate we can set $B_0=1$. 
Notice next that the coordinate $x$ can be compact or not. Below we discuss these two possibilities.\\
\textbf{\underline{Non-compact $x$ coordinate:}}\\
In what follows we will take $B_0=1$. Take the range of $x$ to be $(-\infty,\infty)$ and impose the boundary condition that the boundary metric (i.e., at $r\rightarrow\infty$) is asymptotically flat. Then for the region $x\geq 0$ we must set $A_2=0$ while for the region $x\leq 0$ we must set $A_1=0$. We also want to require that the solution is symmetric with respect to $x=0$. Altogether, the solution will read,
\beq
ds^2=\frac{L^2dr^2}{r^2\left(1-ML^2/r^2\right)}+\frac{r^2}{L^2}\left[-\left(1-ML^2/r^2\right)\left(1+\frac{Ae^{\mp\sqrt{M}x/L}}{\sqrt{1-ML^2/r^2}}\right)^2dt^2+dx^2\right]\,.
\eeq
It is appropriate to write the constant $A$ as $A=-e^{\sqrt{M}x_0/L}$ where $x=\pm x_0$ are the locations of the boundary horizon (as can be easily checked). Thus we finally have,
\beq
ds^2=\frac{L^2dr^2}{r^2\left(1-ML^2/r^2\right)}+\frac{r^2}{L^2}\left[-\left(1-ML^2/r^2\right)\left(1-\frac{e^{\mp\sqrt{M}(x\mp x_0)/L}}{\sqrt{1-ML^2/r^2}}\right)^2dt^2+dx^2\right]\,,
\eeq
which, as discussed in section \ref{subsec:Black Funnels}, is the metric of a black funnel. See Fig.[\ref{fig:FIG3}] for the shape of the horizon.\\
\textbf{\underline{Compact $x$ coordinate:}}\\
If, on the other hand, we take the $x$ coordinate to be compact, it is appropriate to define the angle $\theta=x/L$ where $\theta \in[0,2\pi]$. Then we have,   
\beq
ds^2=-\left(r^2/L^2-M\right)\left(1+\frac{A_1 e^{-\sqrt{M}\theta}+A_2 e^{\sqrt{M}\theta}}{\sqrt{1-ML^2/r^2}}\right)^2dt^2+\frac{dr^2}{r^2/L^2-M}+r^2d\theta^2\,.
\eeq   
Now we must require (or make sure) that the solution is periodic in $\theta$. There is more than one way how to make the solution periodic, and we are going to choose one. We are going to choose a solution which is symmetric around $\theta=0$, namely, we will take $A_1=A$, $A_2=0$ for $\theta \in[0,\pi]$, and $A_1=0$, $A_2=A$ for $\theta \in[-\pi,0]$. The solution will read,  

\beq
ds^2=-\left(r^2/L^2-M\right)\left(1+\frac{Ae^{\mp\sqrt{M}\theta}}{\sqrt{1-ML^2/r^2}}\right)^2dt^2+\frac{dr^2}{r^2/L^2-M}+r^2d\theta^2\,,
\eeq
where the upper sign refers to the range $0\leq\theta\leq\pi$ while the lower sign refers to the range $-\pi\leq\theta\leq 0$. It is appropriate to write $A$ as $A=-e^{\sqrt{M}\theta_0}$, and thus we finally have,

\beq
ds^2=-\left(r^2/L^2-M\right)\left(1-\frac{e^{\mp\sqrt{M}(\theta\mp\theta_0)}}{\sqrt{1-ML^2/r^2}}\right)^2dt^2+\frac{dr^2}{r^2/L^2-M}+r^2d\theta^2\,,
\eeq
where $\theta=\pm\theta_0$ are the locations of the boundary horizon as discussed in section \ref{sec:General Solution}.

\section{General Static Metric in Three Dimensions}\label{app:gen}
\underline{\textbf{Claim:}} The most general static metric in $3$ dimensions can be written as,
\beq\label{met}
ds^2=g_{rr}(r,x)dr^2+g_{tt}(r,x)dt^2+g_{xx}(r,x)dx^2\,,
\eeq
where $g_{rr}(r,x)$ is an arbitrary function.\\
\underline{\textbf{Proof:}} Start from the general form,
\beq
ds^2=g_{rr}(r,x)dr^2+2g_{rt}(r,x)drdt+2g_{rx}(r,x)drdx+g_{tt}(r,x)dt^2+2g_{tx}(r,x)dtdx+g_{xx}(r,x)dx^2\,.
\eeq
First, since we are looking for non-rotating solutions we set,
\beq
g_{rt}=g_{tx}=0\,.
\eeq 
Second, it can be easily checked by a straightforward calculation that by a coordinate transformation of the form $r=A(r',x')$, $x=B(r',x')$ we can make 
\beq
g_{r'x'}=0\,,
\eeq 
and 
\beq
g_{r'r'}=\text{arbitrary}\,,
\eeq 
by choosing the functions $A$ and $B$ appropriately. Hence, the metric reduces to,  
\beq
ds^2=g_{r'r'}(r',x')dr'^2+g_{tt}(r',x')dt^2+g_{x'x'}(r',x')dx'^2\,,
\eeq
with $g_{r'r'}$ being arbitrary. By removing the primes over $r'$ and $x'$ we obtain (\ref{met}) as claimed.

\section{The General Solution (Derived from the Case [2])}\label{app:bbb}

As mentioned before in this case $B'(x)\neq 0$ and $G(x)=F(x)A'(x)/B'(x)$. Thus, the metric will read [see (\ref{general metric}),(\ref{g}),(\ref{f})],

\begin{align}\label{met1b}
ds^2&=\frac{L^2dr^2}{r^2\left(1-ML^2/r^2\right)}+\frac{r^2}{L^2}\left[-\left(1-ML^2/r^2\right)\left(\frac{A(x)}{\sqrt{1-ML^2/r^2}}+B(x)\right)^2dt^2 \right.\nonumber\\
&\qquad \left.+F(x)^2\left(\sqrt{1-ML^2/r^2}+\frac{A'}{B'}\right)^2dx^2\right]\,.
\end{align}
There is still one equation to satisfy, namely, the $rr$ component (\ref{eom1b}), which we repeat here for convenience, 
\beq\label{eom1b'}
MF^3AA'+B'\left[-MF^3B+L^2\left(B'F'-FB''\right)\right]=0\,. 
\eeq
It is obvious that the function $F(x)$ can not be identically zero, since otherwise $g_{xx}=0$ [see (\ref{met1b})] and so this metric will be ruled out then.
Since $F\neq0$ we can by a reparameterization of the $x$ coordinate set $F=1$. Upon the last step the $rr$ equation (\ref{eom1b'}) simplifies and reduces to, 
\beq\label{eom1b''}
AA'=B'\left[B+\frac{L^2}{M}B''\right]\,. 
\eeq
Before solving this equation let us fix some gauge freedom for the sake of simplicity. Look at the boundary metric obtained by taking the limit $r\rightarrow\infty$, 
\begin{equation}
ds_{1+1}^2=-\left(A+B\right)^2dt^2+\left(1+\frac{A'}{B'}\right)^2dx^2\,,
\end{equation}
and without loss of generality require that it takes the form (recall that any $1+1$ metric can be put in the following form by a coordinate transformation)
\begin{equation}
ds_{1+1}^2=-f(x)^2dt^2+dx^2\,.    
\end{equation}
Thereafter, we get a set of three equations to solve, $\left(A+B\right)^2=f^2$ and $\left(1+A'/B'\right)^2=1$, in addition to $AA'=B'\left[B+\frac{L^2}{M}B''\right]$. Solving these three equations yields two metrics, one of which is already known (found) to us while the other looks unfamiliar (new). The new metric reads,
\begin{align}\label{}
ds^2&=\frac{L^2dr^2}{r^2\left(1-ML^2/r^2\right)}+\frac{r^2}{L^2}\left[-\left(1-ML^2/r^2\right)\left(\frac{2f(x)-3C_0}{\sqrt{1-ML^2/r^2}}+3C_0-f(x)\right)^2dt^2 \right.\nonumber\\
&\qquad \left.+\left(2-\sqrt{1-ML^2/r^2}\right)^2dx^2\right]\,,
\end{align}
where 
\beq
f(x)=C_0+C_1 e^{-\sqrt{3M}x/L}+C_2 e^{\sqrt{3M}x/L}\,,
\eeq
where $C_0$, $C_1$ and $C_2$ are arbitrary constants. For reasons to be explained momentarily we will perform the change $M\rightarrow M/3$ and so,
\begin{align}\label{}
ds^2&=\frac{L^2dr^2}{r^2\left(1-ML^2/3r^2\right)}+\frac{r^2}{L^2}\left[-\left(1-ML^2/3r^2\right)\left(\frac{2f(x)-3C_0}{\sqrt{1-ML^2/3r^2}}+3C_0-f(x)\right)^2dt^2 \right.\nonumber\\
&\qquad \left.+\left(2-\sqrt{1-ML^2/3r^2}\right)^2dx^2\right]\,,
\end{align}
where 
\beq
f(x)=C_0+C_1 e^{-\sqrt{M}x/L}+C_2 e^{\sqrt{M}x/L}\,.
\eeq

We can compactify the $x$ coordinate by defining the angle $\theta=x/L$ with the range $[0,2\pi]$. The solution reads,
\begin{align}\label{case[2]}
ds^2&=-\left(r^2/L^2-M/3\right)\left(\frac{2f(\theta)-3C_0}{\sqrt{1-ML^2/3r^2}}+3C_0-f(\theta)\right)^2dt^2+\frac{dr^2}{r^2/L^2-M/3} 
\nonumber\\
&\qquad +r^2\left(2-\sqrt{1-ML^2/3r^2}\right)^2d\theta^2\,,
\end{align}
where 
\beq
f(\theta)=C_0+C_1 e^{-\sqrt{M}\theta}+C_2 e^{\sqrt{M}\theta}\,.
\eeq

In fact, even though the solutions obtained from the case [1b] and case [2] (namely (\ref{general}) and (\ref{case[2]}) respectively) look to the first sight as different they are physically the same. We have concluded that they are the same solution not by finding the coordinate transformation which connects them but by checking that the two solutions have the same boundary metric and the same boundary stress tensor; recall that, in gravity, given the induced metric $\gamma_{ab}$ and the extrinsic curvature $\Theta_{ab}$ (or equivalently $T_{ab}$) on a Cauchy surface $\Sigma$ then a unique solution is gauranteed.

\end{document}